\documentclass[
    ,final            
  ]
  {aipproc}

\layoutstyle{6x9}

\begin{document}

\title{Extraction of polarized parton densities from polarized DIS and SIDIS.}

\classification{12.38.Bx,13.85.Ni}
\keywords      {polarized PDF, DIS, SIDIS}

\author{D. de Florian}{ address={Departamento de F\'{\i}sica,
Universidad de Buenos Aires, Ciudad Universitaria, Pab.1 (1428)
Buenos Aires, Argentina}}

\author{G. A Navarro}{ address={Departamento de F\'{\i}sica,
Universidad de Buenos Aires, Ciudad Universitaria, Pab.1 (1428)
Buenos Aires, Argentina}}

\author{R. Sassot}{
  address={Departamento de F\'{\i}sica,
Universidad de Buenos Aires, Ciudad Universitaria, Pab.1 (1428)
Buenos Aires, Argentina}}

\begin{abstract}
We present results on the quark and gluon polarization in the nucleon 
obtained in a combined next to leading order analysis to the available 
inclusive and semi-inclusive polarized deep inelastic scattering data.
\end{abstract}

\maketitle


\section{Introduction}
For more than fifteen years, polarized inclusive deep inelastic scattering 
(pDIS) has been the main source of information on how the 
individual partons in the nucleon are polarized at very short distances. 
Many alternative experiments have been conceived to improve this situation. 
The most mature among them are those based on polarized semi-inclusive deep 
inelastic scattering (pSIDIS).

In the following we present results obtained in a combined next to leading 
order analysis to the recently updated set of pDIS and pSIDIS data 
\cite{deFlorian:2005mw}.
Specifically, we focused on the extraction of sea quark and gluon densities, 
analyzing the constraining power of the data on the individual densities. 
As result, we found not only a complete agreement between pDIS and pSIDIS 
data, but a very useful complementarity, leading to rather well 
constrained densities.   

Using the Lagrange multiplier approach \cite{Stump:2001gu}, we explored the 
profile of the $\chi ^2$ function against different degrees of polarization 
in each parton flavor. In this way we obtained estimates for the uncertainty 
in the net polarization of each flavor, and in the parameters of the pPDFs.  
We compared results obtained with the two most recent sets of fragmentation 
functions. The differences are found to be within conservative estimates for 
the uncertainties. Nevertheless, there is a clear preference for a given set 
of FF over the other, shown in a difference of several units in the $\chi ^2$ 
of the respective global fits. In NLO global fits the overall agreement 
between theory and the full set of data is sensibly higher than in LO case.

\section{Global Fit}

In our analysis \cite{deFlorian:2005mw}, we followed the same conventions 
and definitions for the polarized inclusive asymmetries and parton densities 
adopted in references \cite{DS,NS}, however we used more recent inputs, such 
as unpolarized parton densities \cite{MRST02} and the respective values for 
$\alpha_s$. Fragmentation functions were taken from either \cite{kretzer} or 
\cite{KKP}, respectively. We also used the flavor symmetry and flavor 
separation criteria proposed in \cite{kretzer}, at the respective initial 
scales $Q^2_{i}$. The data sets analyzed  include only points with $Q^2>1$
GeV$^2$, listed in Table \ref{tab:table1}, and totaling 478 data points.
 
\begin{table}
\begin{tabular}{ccccc} \hline
Collaboration & Target& Final state & \# points & Refs. \\ \hline
EMC          & proton& inclusive   &    10     & \cite{see} \\ 
SMC          & proton, deuteron & inclusive &  12, 12  & \cite{see} \\  
E-143        & proton, deuteron & inclusive  &  82, 82  & \cite{see} \\ 
E-155        & proton, deuteron & inclusive   &    24, 24    & \cite{see} \\ 
Hermes       & proton,deuteron,helium& inclusive   &    9, 9, 9   & \cite{see} \\
E-142        & helium& inclusive   &    8     & \cite{see} \\ 
E-154        & helium& inclusive   &    17     & \cite{see} \\ 
Hall A       & helium & inclusive &   3      & \cite{see}   \\
COMPASS      & deuteron & inclusive &   12  & \cite{see} \\
 \hline 
SMC          & proton,deuteron& $h^+$, $h^-$  &  24, 24  & \cite{see} \\ 
Hermes       & proton, deuteron, helium & $h^+$, $h^-$, $\pi^+$, $\pi^-$, $K^+$, $K^-$, $K^T$   &    36,63,18     
& \cite{see} \\  \hline
\multicolumn{3}{c}{Total} & 478 & \\ \hline
\end{tabular}
\caption{\label{tab:table1} Inclusive and semi-inclusive data used in the fit.}
\end{table}

In Table \ref{tab:table2}, we summarize the results of the best NLO and LO 
global fits to all the data listed in Table \ref{tab:table1}. We present fits 
obtained using alternatively fragmentation functions from 
reference \cite{kretzer}, labeled as  KRE, and from reference \cite{KKP}, 
labeled as KKP.
\begin{table}[b]
\caption{\label{tab:table2} $\chi^2$ values and first moments for distributions at $Q^2=10$ GeV$^2$}
\begin{tabular}{cccccccccc} \hline 
\multicolumn{2}{c}{set}  & $\chi ^2$ &$\chi^2_{DIS}$ &$\chi^2_{SIDIS}$  & $\delta \overline{u}$ & 
$\delta \overline{d}$   & $\delta \overline{s}$ & 
$\delta g$ & $\delta \Sigma$\\ \hline
& KRE & 430.91 &206.01 & 224.90   & -0.0487 &-0.0545 & -0.0508 & 0.680 & 0.284\\ 
\raisebox{1.7ex}{NLO}  & KKP & 436.17&205.66&230.51 & 0.0866 &-0.107 & -0.0454  & 0.574 & 0.311\\ \hline
 & KRE  & 457.54&213.48 &244.06    & -0.0136  &-0.0432  & -0.0415 & 0.121 &
 0.252 \\
\raisebox{1.7ex}{LO}  & KKP  & 448.71&219.72  & 228.99  & 0.0497& -0.0608 & -0.0365 &0.187 & 0.271 \\ \hline
\end{tabular}
\end{table}
Since the fit involves 20 parameters, the number of degrees of freedom for 
these fits is 478-20=458. Consequently, the $\chi^2$ values obtained are 
excellent for NLO fits and very good for LO. The better agreement between 
theory and experiment found at NLO, highlights the importance of the 
corresponding QCD corrections, for the present level of accuracy achieved by 
the data. 

In NLO fits there seems to be better agreement when using KRE 
fragmentation functions.
The difference between the total $\chi^2$ values between KRE and KKP NLO fits 
comes mainly from the contributions related to pSIDIS data, while those 
associated to inclusive data are almost the same, as one should expect in a 
fully consistent scenario.

Table  \ref{tab:table2} includes also the first moment of each flavor 
distribution at $Q^2=10$ GeV$^2$, and that for the singlet distribution 
$\delta \Sigma$, as reference.
Most noticeably, while the KRE NLO fit favors the idea of a SU(3) symmetric 
sea, KKP NLO finds $\overline{u}$ polarized opposite to $\overline{d}$ and to $\overline{s}$. Gluon and strange sea quark polarization are similar in both 
fits and the total polarization carried by quarks is found to be around 30\%.

\section{Uncertainties}
Many strategies have been implemented in order  to assess the uncertainties in 
PDFs and their propagation to observables, specially those associated with experimental errors in the data. The Lagrange multiplier method \cite{Stump:2001gu} 
probes the uncertainty in any observable or quantity of interest relating 
the range of variation of one or more physical observables dependent upon PDFs 
to the variation in the $\chi^2$ used to judge the goodness of the fit to data. \setlength{\unitlength}{1.mm}
\begin{figure}[hbt]
\includegraphics[width=14cm]{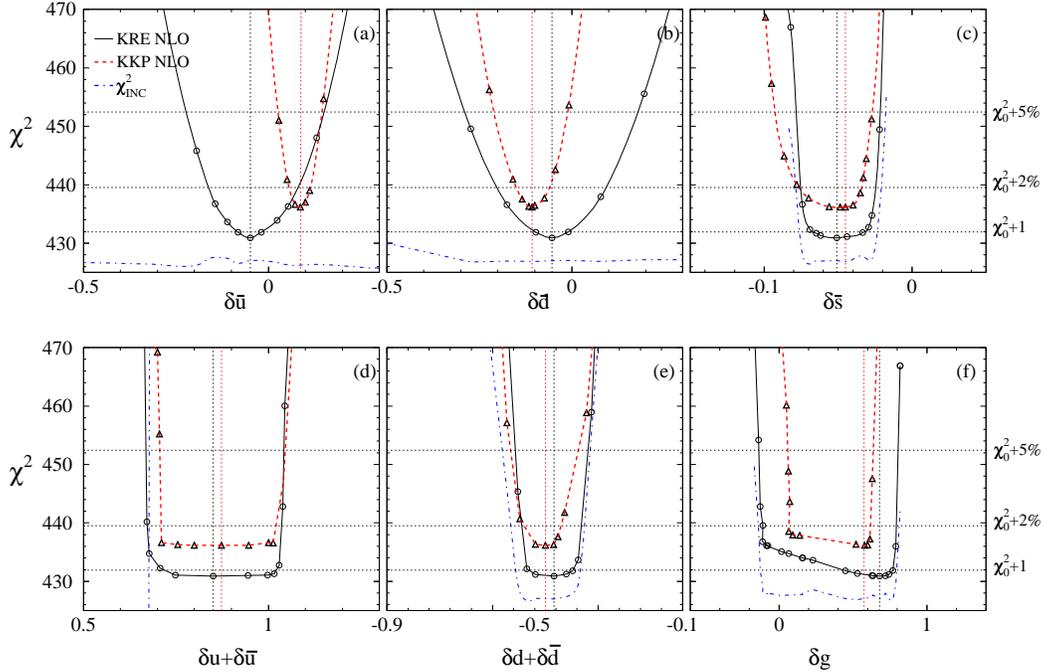}
\caption{$\chi^2$ profiles for NLO fits obtained using Lagrange multipliers
at $Q^2=10$ GeV$^2$.}
\label{fig:krenlo}
\end{figure}
 In Figure \ref{fig:krenlo}  we show the outcome of varying the $\chi^2$ of 
the NLO fits to data against the first moment of the respective polarized 
parton densities $\delta q$ at $Q^2=10$ GeV$^2$, one at a time. This is, to 
minimize
\begin{equation}
\Phi(\lambda_q, a_j)=\chi^2(a_j)+\lambda_q\, \delta q(a_j) \,\,\,\,\,\,\,\,\,\,\,\, q=u,\overline{u},d,\overline{d},s,g.
\end{equation}

\setlength{\unitlength}{1.mm}
\begin{figure}[hbt]
\includegraphics[width=14cm]{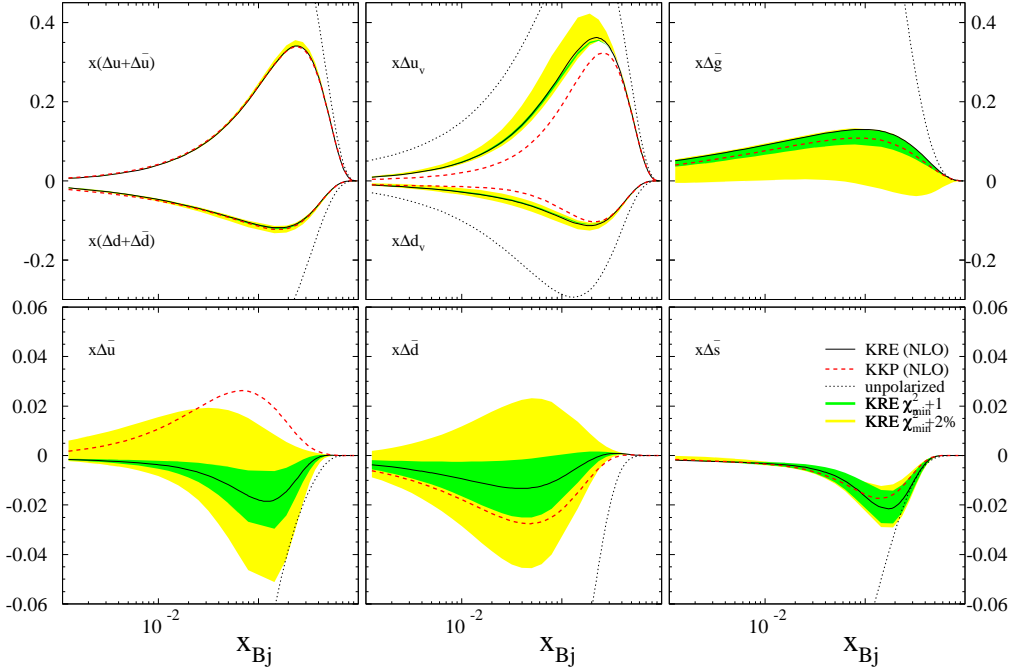}
\caption{Parton densities at $Q^2=10$ GeV$^2$, and the uncertainty bands corresponding to $\Delta \chi^2=1$ and $\Delta \chi^2=2\%$  }
\label{fig:disnlo}
\end{figure}

In order to see the effect of the variation in $\chi^2$ on the parton 
distributions themselves, in Figure  \ref{fig:disnlo}, we show KRE best fit
densities together with the uncertainty bands corresponding to 
$\Delta \chi^2=1$ (darker band) and $\Delta \chi^2=2\%$ (light shaded band). 
As expected, the relative uncertainties in the total quark densities and 
those strange quarks are rather small. For gluon densities the 
$\Delta \chi^2=1$ band is also small, but the most conservative 
$\Delta \chi^2=2\%$ estimate is much more significative. For light sea quarks 
the  $\Delta \chi^2=1$ bands are moderate but the $\Delta \chi^2=2\%$ are much more larger.

Two programmed experiments, the one based on the PHENIX detector already 
running at RHIC \cite{Fukao:2005id}, and the E04-113 experiment at JLab  
\cite{x} will be able to 
reduce dramatically the uncertainty in both the gluon and the light sea 
quark densities respectively, the latter providing also an even more
stringent test on fragmentation functions.

\begin{theacknowledgments}
Partially supported by CONICET, Fundaci\'on Antorchas, UBACYT and 
ANPCyT, Argentina.
\end{theacknowledgments}




\end{document}